\documentclass{appolb}
\usepackage{epsfig}
\def\ee{e^+e^-}
\def\as{\alpha_s}
\def\Ko{K_{\rm out}}
\def\cO#1{{\cal{O}}\left(#1\right)}
\def\cP{{\cal {P}}}
\begin{document}
\pagestyle{plain}
\newcount\eLiNe\eLiNe=\inputlineno\advance\eLiNe by -1
\title{OUT-OF-PLANE QCD RADIATION \\
IN DIS EVENTS WITH HIGH $P_T$ JETS 
\thanks{Talk presented at the DIS 2002 conference.}}%
\author{Giulia ZANDERIGHI
\address{Institute of Particle Physics Phenomenology, University of Durham, \\Durham DH1 3LE, England.}}
\maketitle

\begin{abstract}
We present the QCD analysis of the out-of-event-plane momentum
distribution in DIS events with high $p_t$ jets. The achieved accuracy
allows the measurement of the running coupling and the study of
non-perturbative effects, in particular the test of universality of
power corrections in a new experimental regime.
\end{abstract}

\section{Introduction}
Event shape variables describe the energy and momentum flow in high
energy collisions.
Being collinear and infrared safe they can be computed with high
accuracy in QCD so that they generally allow a good measurement of
$\alpha_s$.
Furthermore they are sensitive to extra soft non perturbative
emissions, so that they prove to be a powerful tool to study the up to
now fairly unknown low energy domain.
\section{Two- and multi-jet event shape variables}
Event shapes were originally defined for $\ee$
collisions~\cite{CTTW,broad,NewBroad}, only quite recently their QCD
calculation have been extended to Deep Inelastic Scattering (DIS)
collisions~\cite{TDIS,BDIS}.
Calculations for DIS processes are in principle a simple extension of
the $\ee$ case. Of course just from a pure kinematical point of view
one expects some differences: $\ee$ collisions are characterized by
only one hard scale, the center-of-mass energy $\sqrt{s}$, while DIS
processes depend on one additional hard scale, the virtuality
$Q$ of the incoming gauge boson.
A further difference is the presence of a hadron in the initial state
in DIS collisions. This implies that observables are finite only {\em
after} introducing a factorization scale to extract collinear
divergences due to initial state radiation.  Finally, since no
measurements can be performed close to the beam region, in DIS
analyses one needs to introduce a kinematical rapidity cut
$\eta<\eta_0$ ($\sim\cO{1}$) along the direction of the incoming
parton.

Till recently, both in $\ee$ and in DIS physics, attention was focused
on so called 2-jet observables, i.e. on those observables whose first
non-zero contribution is at order $\as$. Such observables are
sensitive to any kinematical final state, therefore prove to have a
rich phenomenology~[6 - 13]. 
On the other hand, 3-jet observables, i.e. observables whose first
non-vanishing contribution is at order $\as^2$ are sensitive to the
non-abelian structure of QCD already at leading order and are
therefore more interesting from a theoretical point of view.
In particular in DIS processes 3-jet observables are sensitive to the
gluon density at leading order.
Therefore these studies should provide a powerful method to 
study QCD dynamics and to constrain the parton distribution
functions. 
\section{Out-of-plane radiation}
The computation of 3-jet observables at the same standard accuracy
available for 2 jet observables (i.e. single logarithmic resummation,
second order exact results, matching and leading power corrections)
proves to be much more cumbersome.
This project started some years ago with the study of some 3-jet
observables in $\ee$~\cite{Tmin,Dpar}, results were then
extended to Drell Yan processes~\cite{kouthh} and finally to DIS
observables~\cite{KoutDIS,Azcorr}.
The first $3$-jet observable studied in DIS measures the
cumulative out-of-plane momentum distribution $\Ko$~\cite{KoutDIS}, where
the event-plane is fixed kinematically by the Breit- and thrust major
axis.
Pure three jet events are selected by imposing a lower bound for the
2-jet resolution variable $y_2$.

At first order in $\as$, i.e. when only 2 final state partons are
present, momentum conservation ensures that the event is planar so
that the observable vanishes. The first non trivial contribution is
therefore at $\cO{\as^2}$. 

The $\Ko$ observable, as other event shapes, is characterized by
the following theoretical features:
\begin{itemize}
\item 
on a perturbative (PT) level, in the kinematical region where \mbox{$\Ko\ll Q$}
vetoing real emissions gives rise to a large mismatch between real and
virtual contributions, so that large logarithms $L\equiv\ln(\Ko/Q)$ need
to be resummed at double- (DL) and single logarithmic (SL)
level.
\item
from a non perturbative (NP) point of view also ultra-soft
emissions contribute to the observable, so that (at least) leading $1/Q$ power
corrections need to be taken into account.
\end{itemize}

\subsection{Perturbative result}
At DL level the perturbative answer is quite straightforward: vetoing
radiation from any of the 3 emitting partons gives rise to a Sudakov
exponent which is simply the sum of the 3 single-parton Sudakovs, each
one forbidding radiation from one hard parton regardless of the
presence of the other 2.
The Sudakov factor needs to be weighted with the hard squared matrix
element $|M|^2$ and convoluted with the incoming parton density
$\cP(\mu)$
\begin{equation}
\Sigma(\Ko) \sim \cP(\mu)\otimes
|M|^2\cdot e^{-\frac{\as}{\pi}(2C_F+C_A)\ln^2\frac{Q}{\Ko}}\>. 
\end{equation}

A naive DL result is known to be insufficient to make quantitative
predictions.  A more rich and informative answer is found at SL level
where the answer is sensitive to coherence and interference effects.
In particular to hard parton recoil effects, to soft large angle
radiation, to collinear (initial) state radiation and to the running
of $\as$ (at 3 different scales).

The total answer at SL level~\cite{KoutDIS} turns out to be quite
involved, but can be interpreted easily in terms of the above SL
effects.
\subsection{Non-perturbative effects}
As is now well established, for event shapes PT results at whatever
order need to be supplemented by an NP contribution whose actual size
and form depends strictly on what has been already included in the PT
calculations.
The kinematical origin of such a sensitivity to hadronization effects
is evident: event shapes measure momentum flow, they are therefore
sensitive to the momentum degradation which occurs during the colour
blanching.
Also from a more formal point of view the need for NP corrections is
clear: 
regardless of details of the behaviour of the coupling in the infrared
PT expansions are divergent and therefore intrinsically ambiguous,
such an ambiguity in the final answer needs to be canceled by an
additional NP term.

To deal with NP effects we adopted the now standard procedure of
extending the notion of $\as$ in the NP regime~\cite{DMW}.  The total
answer depends then on $\as$ and on one NP parameter $\alpha_0$ which
is related to the average of the dispersive coupling in the NP domain
(the choice of
$\mu_I$ 
is conventional)
\begin{equation}
\alpha_0=\frac{1}{\mu_I}\int_0^{\mu_I} dk_t\, \as(k_t)\>,
\qquad \mu_I \sim 2\, \mbox{GeV}\>.
\end{equation}
$\alpha_0$ is the same parameter which appears in the calculation of
other event shapes. The fact that many observables depend on
only one additional NP parameter is a consequence of the universal
(linear) behaviour of such observables on the transverse momentum of the
secondary partons.
The rapidity (and azimuthal) dependence determines the
observable-dependent, PT-calculable coefficient of the NP corrections.
In the present case, since $\Ko$, as the Broadenings, is by definition
rapidity-independent, the rapidity integral of the NP gluons would
diverge if extended naively to infinity as is usually done. One then
needs to introduce an effective rapidity cutoff, which is provided by
the PT recoil of the hard partons.
Due to this interplay between PT \& NP physics the NP-shift depends on
the PT value of the observable. In particular, according to the
kinematical region under consideration the shift is log-enhanced
($\propto \ln (\Ko/{Q})$) or proportional to $1/\sqrt{\as}$ (with a
rich and informative colour structure).
\subsection{Universality of power corrections}
\begin{figure}
\begin{center}
\epsfig{file=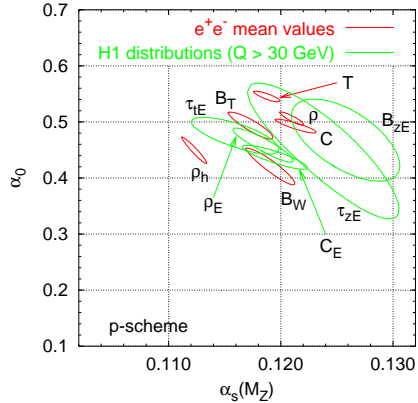,width=0.43\textwidth}
    \caption{1-$\sigma$ contour plot for the $\as-\alpha_0$ fit for
    some DIS event shape variables.}  \label{fig:asa0}
\end{center}
\end{figure}

Universality implies that a simultaneous fit of $\as$ and $\alpha_0$
from different observables gives the same answer.  Fig.~\ref{fig:asa0}
shows a 1-$\sigma$ contour plot for some DIS event
shapes~\cite{ref:asa0}.
Universality is found with an accuracy of about \mbox{$10-15\%$},
which is in agreement with the size of higher order neglected
terms. Up to now
this universality hypothesis has been tested only in 2-jet
observables, so that there is a quite large correlation between
different fits which is usually not taken into account.
Indeed fits are done using the same experimental data and fitting
observables which are fairly similar (think of the thrust $T$ and
$C$-parameter for which at leading order $C\,=\,6\, (1\,-\,T)$).

The computation of the out-of-plane momentum distribution provides a
possibility of checking this universality hypothesis in a more
uncorrelated environment.  This check is particularly relevant since
universality of $\alpha_0$ relies on the assumption of NP gluons being
distributed uniformally in rapidity, which is quite non trivial in the
3-jet case.
\section*{Acknowledgments}
This work has been done in collaboration with Andrea Banfi, Pino
Marchesini and Graham Smye.
Thanks also to Yuri Dokshitzer and Gavin Salam for many precious
hints.

\end{document}